\documentclass[fleqn,usenatbib]{mnras} %\onecolumn

\usepackage{amsmath}	% Advanced maths commands
\usepackage{amssymb}	% Extra maths symbols
\usepackage{tikz}

\usepackage{lipsum}                  % Para generar texto falso.
\usepackage[normalem]{ulem}          % Permite tachar texto

\usepackage{graphicx}

\title[mHz QPO frequency drift in 4U~1608--52 and Aql X--1]{Drifts of the marginally stable burning frequency in the X-ray binaries 4U~1608--52 and Aql X--1}

\author[G. C. Mancuso et al. ]{G. C. Mancuso,$^{1,2}$\thanks{E-mail: gmancuso@iar.unlp.edu.ar} D. Altamirano,$^{3}$ M. M\'{e}ndez,$^{4}$ M. Lyu,$^{5}$ and J. A. Combi$^{1,2}$\\
$^{1}$Instituto Argentino de Radioastronom\'{\i}a (CCT-La Plata, CONICET; CICPBA), C.C. No. 5, 1894 Villa Elisa, Argentina\\
$^{2}$Facultad de Ciencias Astron\'omicas y Geof\'{\i}sicas, Universidad Nacional de La Plata, Paseo del Bosque s/n, 1900 La Plata, Argentina\\
$^{3}$Physics \& Astronomy, University of Southampton, Southampton, Hampshire SO17 1BJ, UK\\
$^{4}$Kapteyn Astronomical Institute, University of Groningen, PO BOX 800, NL-9700 AV Groningen, the Netherlands\\
$^{5}$Department of Physics, Xiangtan University, Xiangtan, Hunan 411105, China}

\begin{document}
%
%
%\pagerange{\pageref{firstpage}--\pageref{lastpage}} \pubyear{2014}
%
\maketitle
\label{firstpage}
\begin{abstract}
We detect millihertz quasi-periodic oscillations (mHz QPOs) using the Rossi X-ray Time Explorer (RXTE)
from the atoll neutron-star (NS) low-mass X-ray binaries 4U~1608--52 and Aql X--1. 
From the analysis of all RXTE observations of 4U~1608--52 and Aql X--1, we find mHz QPOs with a significance level $>3\sigma$ in 49 and 47 observations, respectively. 
The QPO frequency is constrained between $\sim$ 4.2 and 13.4 mHz. 
These types of mHz QPOs have been interpreted as being the result of marginally stable nuclear burning of He on the NS surface.
We also report the discovery of a downward frequency drift in three observations of 4U~1608--52, making it the third source that shows this behaviour. 
We only find strong evidence of frequency drift in one occasion in Aql X--1, probably because the observations were too short to measure a significant drift. 
Finally, the mHz QPOs are mainly detected when both sources are in the soft or intermediate states; the cases that show frequency drift only occur when the sources are in intermediate states. 
Our results are consistent with the phenomenology observed for the NS systems 4U~1636--53 and EXO~0748--676, suggesting that all four sources can reach the conditions for marginally stable burning of He on the NS surface. 
These conditions depend on the source state in the same manner in all four systems.\\

\end{abstract}


%%%%%%%%%%%%%%%%%%%%%%%%%%%%%%%%%%%%%%%%%%%%%%%%%%%%%%%%%%%%%%%%%%%%%%%%%%%%%%%
%%%%%%%%%%%%%%%%%%%%%%%%%%%%%%%%%%%%%%%%%%%%%%%%%%%%%%%%%%%%%%%%%%%%%%%%%%%%%%%
%%%%%%%%%%%%%%%%%%%%%%%%%%%%%%%%%%%%%%%%%%%%%%%%%%%%%%%%%%%%%%%%%%%%%%%%%%%%%%%
%
%
% Introduction
%
%
%%%%%%%%%%%%%%%%%%%%%%%%%%%%%%%%%%%%%%%%%%%%%%%%%%%%%%%%%%%%%%%%%%%%%%%%%%%%%%%
%%%%%%%%%%%%%%%%%%%%%%%%%%%%%%%%%%%%%%%%%%%%%%%%%%%%%%%%%%%%%%%%%%%%%%%%%%%%%%%
%%%%%%%%%%%%%%%%%%%%%%%%%%%%%%%%%%%%%%%%%%%%%%%%%%%%%%%%%%%%%%%%%%%%%%%%%%%%%%%

\section{Introduction}\label{sec:intro}

Low-mass X-ray binaries (LMXBs) are systems harbouring a compact object, either a neutron star (NS) or a black hole, and a low-mass ($\lesssim$~1\hspace{.7mm}$M_{\odot}$) donor star. The compact object accretes material from the companion star through an accretion disc (\citealt{pringle1972}; \citealt{shakura1973}). These LMXBs can be always active, showing a persistent X-ray luminosity in the 10$^{36-38}$ erg s$^{-1}$ range, for years to decades, or transient, spending most of the time in a low luminosity or quiescence state ($L_{\rm X} \lesssim 10^{34}$ erg s$^{-1}$; see, e.g., \citealt{chen1997}). This quiescence state is occasionally interrupted by outbursts when the X-ray luminosity increases by up to several orders of magnitude ($L_{\rm X} \simeq 10^{35-39}$ erg s$^{-1}$; see, e.g., \citealt{lasota2001} for a review).

Bright NS LMXBs are divided into two subclasses, based on their spectral and timing features: atoll sources, with low luminosities ($\lesssim$ 0.5 $L_{\rm Edd}$) and Z sources, with luminosities near (or even above) the Eddington limit (\citealt{hasinger1989}; \citealt{homan2010}). Different spectral states were identified within each of these groups. In particular, the main states of the atoll sources are the extreme island, the island, and the banana branches, a.k.a. hard, intermediate (or transitional), and soft states, respectively (e.g., \citealt{vanderklis2006} and reference therein).

Many NS LMXBs show second to millisecond X-ray variability in their light curves. This variability can be decomposed into separate components in the Fourier power density spectrum (e.g., \citealt{vanderKlis1989}). Depending on the characteristics, they are usually called broad-band noise (with frequencies $<$ 100 Hz) and quasi-periodic oscillations (QPOs; see, e.g., \citealt{vanderklis2006} for a review). QPO appear in a power density spectrum as peaks of finite width which are generally well described by a Lorentzian \citep{belloni2002}. The QPOs display a wide range of frequencies, from millihertz (mHz) to kilohertz \cite[kHz, with frequencies in the 300$-$1300 Hz range;][]{vanderklis2006}. In particular, mHz QPOs were first reported by \citet{revnivtsev2001} in three atoll sources, namely, 4U~1636--53, 4U~1608--52 and Aql X--1, and these QPOs were later detected in three other sources, 4U~1323--619 \citep{strohmayer2011}, GS 1826--238 \citep{strohmayer2018}, and EXO 0748--676 \citep{Mancuso2019}. The mHz QPOs have frequencies between $\sim$ 5 and $\sim$ 15 mHz, and show properties that make them different from all other QPOs seen in NSs: (i) their fractional rms amplitudes are a few percent, increasing with energy from $\sim$ 0.2 to $\sim$ 2.5 keV and then decreasing from $\sim$ 2.5 keV on, becoming undetectable above $\sim$ 5 keV; (ii) they are detected within a narrow range of X-ray luminosities, $L_{2-20\hspace{0.7mm} \rm keV} \simeq (0.5-1.5) \times 10^{37}$ erg s$^{-1}$; and (iii) they are affected by the presence of a thermonuclear (type I) X-ray burst, disappearing at the onset of the burst (\citealt{revnivtsev2001}; \citealt{altamirano2008b}; \citealt{Mancuso2019}; \citealt{lyu2020}); furthermore, there is a strong correlation between the properties of these bursts and mHz QPOs (\citealt{lyu2016}). Although they could not rule out an explanation due to disc instabilities, \citet{revnivtsev2001} related the mHz QPOs with a special regime of nuclear burning on the NS surface. This interpretation was strengthened by the findings of \citet{Yu2002} who reported an anti-correlation between the kHz QPO frequency and the 2--5 keV X-ray count rate associated with a 7.5 mHz QPO in 4U~1608--52. This result is consistent with the inner disc being pushed outward due to the increasing flux during each mHz QPO cycle.

In addition to the six sources mentioned above, there are two other sources with mHz QPOs, but whose features are different compared with those of the previous systems. On one hand, \citet{linares2010} detected QPOs in the mHz range from the NS and 11 Hz pulsar IGR J17480--2446 \citep{strohmayer2010}. The mHz QPOs in this source were found at a persistent luminosity $L_{2-50\hspace{0.7mm} \rm keV} \sim 10^{38}$ erg s$^{-1}$, i.e., roughly an order of magnitude higher than in the other sources, and the QPO frequency was in the range 2.8--4.5 mHz \citep{linares2010,linares2012}. \citet{linares2012} also discovered that the thermonuclear X-ray bursts smoothly evolved into mHz QPOs as accretion rate increased, and vice versa. This is a distinct behaviour compared with the above-mentioned six sources, where mHz QPOs and bursts can be seen within an observation. On the other hand, \citet{ferrigno2017} reported an $\sim$ 8 mHz QPO from the accreting millisecond X-ray pulsar IGR J00291+5934. They found that the QPO was only detected below $\sim$ 3 keV and, although a few scenarios were proposed, the origin of this QPO is still unclear.

The frequency of the mHz QPOs can drift in time. In the case of 4U 1636--53, the mHz QPO sometimes exhibited a continuous decrease of its frequency, and once it dropped below 9 mHz, the oscillations disappeared within a few kiloseconds at the onset of a type I X-ray burst \citep{altamirano2008b}. For EXO 0748--676 a similar drift behaviour was observed in two cases; in one of them, the disappearance of the mHz QPOs at the onset of an X-ray burst was observed \citep{Mancuso2019}.

\citet[][but see \citealt{paczynski1983}]{heger2007} used numerical simulations and reproduced the observed period of the QPO ($\approx$ 100 sec). They also found that the oscillations are only expected in a small range of mass accretion rates, what explains the narrow range of X-ray luminosities where the mHz QPOs are observed. \citet{heger2007} proposed that the physical mechanism behind this oscillatory mode of burning is the marginally stable nuclear burning of He on the surface of an NS. This mode of burning takes place near the boundary between the non-stable and stable burning, where the temperature dependence of the nuclear heating and cooling rates almost cancel.

Although \citet{heger2007} simulations explain the main characteristics of the mHz QPOs, some discrepancies between observations and models are still present. Whereas \citet{heger2007} model predicts that the mHz QPOs should be seen at the transition from unstable to stable burning at practically the Eddington mass accretion rate, $\dot{M}_{\rm Edd}$, mHz QPOs are observed at $\sim 10\%$ $\dot{M}_{\rm Edd}$. In order to explain this difference, \citet{heger2007} proposed that the accreting material would just cover a tenth of the NS surface, so that the local accretion rate is of the order of the Eddington rate. This explanation was reinforced by the results of \citet{lyu2016} who found that in 4U~1636--53 all the type I X-ray bursts that started immediately after mHz QPOs had positive convexities. This, in turn, is related with ignition site of bursts at the NS equator, where the mass accretion rate per unit area might be higher compared with high latitudes. An alternative solution was given by \citet{keek2009}. Using a hydrodynamic stellar evolution code, the authors found that the mixing processes due to rotation (and rotationally induced magnetic fields) combined with a larger energy release from the crust might explain the observed transition boundary. \citet{keek2009} also showed that by lowering the heat flux from the crust, a decrease of the oscillation frequency is seen followed by a flash. This showed that the frequency drift could be the result of the cooling of deep layers as it was proposed by \citet{altamirano2008b}. 

Other major efforts have been made in order to further understand the physical mechanism that produces the mHz QPOs. \citet{lyu2014,lyu2015,lyu2016} did not find any correlation between the persistent flux and frequency evolution nor between the mHz QPO frequency and the temperature of the NS surface in 4U 1636--53. This is at odds with the anti-correlation predicted theoretically \citep{heger2007}. Moreover, \citet{stiele2016} studied the phase-resolved energy spectra of the mHz QPOs in 4U 1636--53 and found that the oscillations were consistent with changes in the size of the emission region and not with variations in the temperature of the NS. 

From the six sources that exhibit mHz QPOs, there are only reports of downward frequency drift in two of them, namely, 4U 1636--53 and EXO 0748--676. \citet{strohmayer2018} analysed all the available observations of GS 1826--238 obtained with the Neutron Star Interior Composition Explorer (NICER) and concluded that the observations were too short to measure potential drifts. They also studied all the observations of GS 1826--238 taken with the Rossi X-ray Timing Explorer (RXTE), but the source was most of the time in the hard state and no mHz QPOs were observed. 4U~1323--619 is a dipper system (\citealt{vanderKlis1985}; \citealt{parmar1989}) so mHz QPOs can potentially be mimicked by regular absorption (dipping) behaviour; therefore a detailed analysis of  4U~1323--619 will be presented elsewhere. In this work we analyse all the RXTE available observations of the remaining two systems: 4U~1608--52 and Aql X--1. 

\vspace{-0.25cm}
\subsection{4U 1608--52}

4U~1608--52 is a transient LMXB discovered in 1972 (\citealt{belian1976}; \citealt{grindlay1976}; \citealt{tananbaum1976}) that undergoes outbursts with a recurrence period ranging from $\sim$ 85 d to $\simeq$ 1--2 yr (\citealt{lochner1994}; \citealt{simon2004}; \citealt{galloway2008}). Type I X-ray bursts were detected by \citet{tananbaum1976}, reveiling the NS nature of the compact object. Burst oscillations in some of the photospheric radius expansion (PRE) bursts were detected at 619 Hz (\citealt{muno2001}; \citealt{galloway2008}). This frequency was associated with the spin period of the NS, making 4U~1608--52 one of the most rapidly rotating accreting NSs \citep{galloway2008}. Single bursts, multi-peak bursts (\citealt{penninx1989}; \citealt{galloway2008}) and a superburst have been observed in this source \citep{keek2008}.

The distance to the source was constrained to the range $\simeq$ 2.9--4.5 kpc (\citealt[][but also see \citealt{guver2010}]{galloway2008,poutanen2014}), and an orbital period of 0.537~d was measured \citep{wachter2002}. \citet{degenaar2015} constrained the disc inclination $i$ to the range $\simeq$ $30\textsuperscript{o}$ -- $40\textsuperscript{o}$. Based on the spectral and timing behaviour, \citet{hasinger1989} classified the system as an atoll source. 

\vspace{-0.25cm}
\subsection{Aql X--1}

Aquila X--1 (hereafter Aql X--1) is a transient LMXB discovered in 1965 by \citet{friedman1967}. Aql X--1 undergoes outbursts quite regularly, with a recurrence time between $\sim$ 125 and $\sim$ 300 d (\citealt{priedhorsky1984}; \citealt{kitamoto1993}; \citealt{campana2013}). \citet{koyama1981} detected type I X-ray bursts, concluding that the compact object in this system is an NS. Similarly to 4U~1608--52,  multi-peaked bursts \citep{galloway2008}, as well as a superburst \citep{serino2016} have been observed from Aql X--1.

Using type I X-ray bursts, \citet{jonker2004} estimated a distance to this system of 4.5--6 kpc, while \citet{matasanchez2017} derived a distance of $d=6 \pm 2$~kpc. Aql X--1 has an orbital period of $\sim$ 18.95 hr (\citealt{chevalier1991}; \citealt{welsh2000}; \citealt{matasanchez2017}), and an inclination angle in the range $36\textsuperscript{o}$ -- $47\textsuperscript{o}$ (\citealt[][but also see \citealt{galloway2016})]{matasanchez2017}. \citet{reig2000,reig2004} studied the colour, spectral and timing properties of Aql X--1, and classified the system as an atoll-type source \citep{hasinger1989}. \citet{casella2008} detected coherent millisecond X-ray pulsations in the persistent X-ray emission at a frequency of 550.27 Hz for $\gtrsim$ 150 sec. This pulsation frequency was slightly higher than the maximum reported frequency from burst oscillations in this source \citep{zhang1998}, and it was associated with the NS spin frequency.

%%%%%%%%%%%%%%%%%%%%%%%%%%%%%%%%%%%%%%%%%%%%%%%%%%%%%%%%%%%%%%%%%%%%%%%%%%%%%%%
%%%%%%%%%%%%%%%%%%%%%%%%%%%%%%%%%%%%%%%%%%%%%%%%%%%%%%%%%%%%%%%%%%%%%%%%%%%%%%%
%%%%%%%%%%%%%%%%%%%%%%%%%%%%%%%%%%%%%%%%%%%%%%%%%%%%%%%%%%%%%%%%%%%%%%%%%%%%%%%
%
%
% Observations and data analysis
%
%
%%%%%%%%%%%%%%%%%%%%%%%%%%%%%%%%%%%%%%%%%%%%%%%%%%%%%%%%%%%%%%%%%%%%%%%%%%%%%%%
%%%%%%%%%%%%%%%%%%%%%%%%%%%%%%%%%%%%%%%%%%%%%%%%%%%%%%%%%%%%%%%%%%%%%%%%%%%%%%%
%%%%%%%%%%%%%%%%%%%%%%%%%%%%%%%%%%%%%%%%%%%%%%%%%%%%%%%%%%%%%%%%%%%%%%%%%%%%%%%

\vspace{-0.25cm}
\section{Observations and data analysis}\label{sec:dataanalysis}

We analysed all public archival observations of both 4U 1608--52 and Aql X--1 taken with the Proportional Counter Array (PCA; \citealt{jahoda2006}) onboard the Rossi X-ray Timing Explorer (RXTE) satellite. We used a total of 1134 and 603 pointed observations of 4U~1608--52 and Aql X--1, respectively, sampling the period comprised between 1996 March and 2011 December. An observation covers one to multiple consecutive RXTE data segments of different lengths separated by data gaps of at least $\sim$ 2.0 ksec. In the case of 4U~1608--52, the data segments last at most $\sim$ 5 ksec, while in the case of Aql X--1, the data segments are not longer than $\sim$ 3.5 ksec. In a handful of cases, the observations were as long as 19.4 ksec. 

We used PCA data obtained in either the Event, Good Xenon, Single Bit or Binned modes. In order to look for mHz QPOs, we applied the Lomb-Scargle periodogram (LSP; \citealt{lomb1976}; \citealt{scargle1982}; \citealt{press1992}) to each individual light curve of 1 sec time resolution to each individual 1 sec light curve in the interval $\approx$ 2--5 keV (generally absolute channels 0--10), where the oscillations are seen to be the strongest \citep{altamirano2008b}. Some observations were performed using the Single Bit mode, which forced us to use the energy ranges $\approx$ 2--6 and 2--10 keV (channels 0--13 and 0--23, respectively). In all those cases in which one or more type I X-ray bursts were observed, we applied the LSP before and after the bursts. We only report all those detections with a significance level $> 3\sigma$.\footnote{In a total of 13 segments (9 of 4U~1608--52 and 4 of Aql X--1), the significance was $\sim$ 3$\sigma$. In order to test whether these detections were significant, we created light curves using the channels 0--8 and computed the LSP again. We found that the significance increased with respect to the value obtained using the channels 0--10, and therefore kept all these segments.}
To calculate the $3\sigma$ level, we followed \citet{press1992} which assumes white noise and takes as number of trials  the number of frequencies searched. To test our assumption on the white noise, we followed \citet{vaughan2005} and fitted a power law to the LSP excluding the QPO. The powerlaw index was consistent with white noise. This is as expected given that the red-noise found in some NS spectral states is the strongest at higher energies than the one we used to search for the mHz QPOs \citep[see, e.g.,][]{vanderklis2006}. For the number of trials, we took the number of frequencies searched in each LSP (which changes as it is dependent on the length of the data-segment; in our case between 500 and 5,000 seconds) so as to take into account the fact QPOs could be detected in a wider range of frequencies. 
We determined the exact mHz QPO frequency, $\nu_{\rm QPO}$, fitting a sinusoidal function to the 1 sec light curve. 

We constructed hardness-intensity and colour-colour diagrams (CCD) of both sources, using the 16-sec time-resolution Standard-2 mode data following the approach described in \citet{Altamirano2005}. In particular, to exclude the X-ray bursts, we removed the data from 10 seconds before to 100 seconds after the onset of each burst. We computed the soft colour as the ratio between the count rates in the 3.5-6.0 keV and 2.0-3.5 keV bands, and the hard colour as the ratio in the 9.7-16.0 keV and 6.0-9.7 keV bands. We defined the intensity as the 2.0-16.0 keV count rate. We normalised all the quantities to those of the Crab in the same energy ranges. Given that both sources transitioned from outburst to quiescence several times, we discarded those observations where the intensity was lower than 5 mCrab.

%%%%%%%%%%%%%%%%%%%%%%% GRAFICO %%%%%%%%%%%%%%%%%%%%%%%%%%%%%%%
\begin{figure*} 
\centering
%\resizebox{2.2\columnwidth}{!}{\rotatebox{0}{\includegraphics[clip]{Figura1-bis-crop.pdf}}}
\resizebox{2.22\columnwidth}{!}{\rotatebox{0}{\includegraphics[clip]{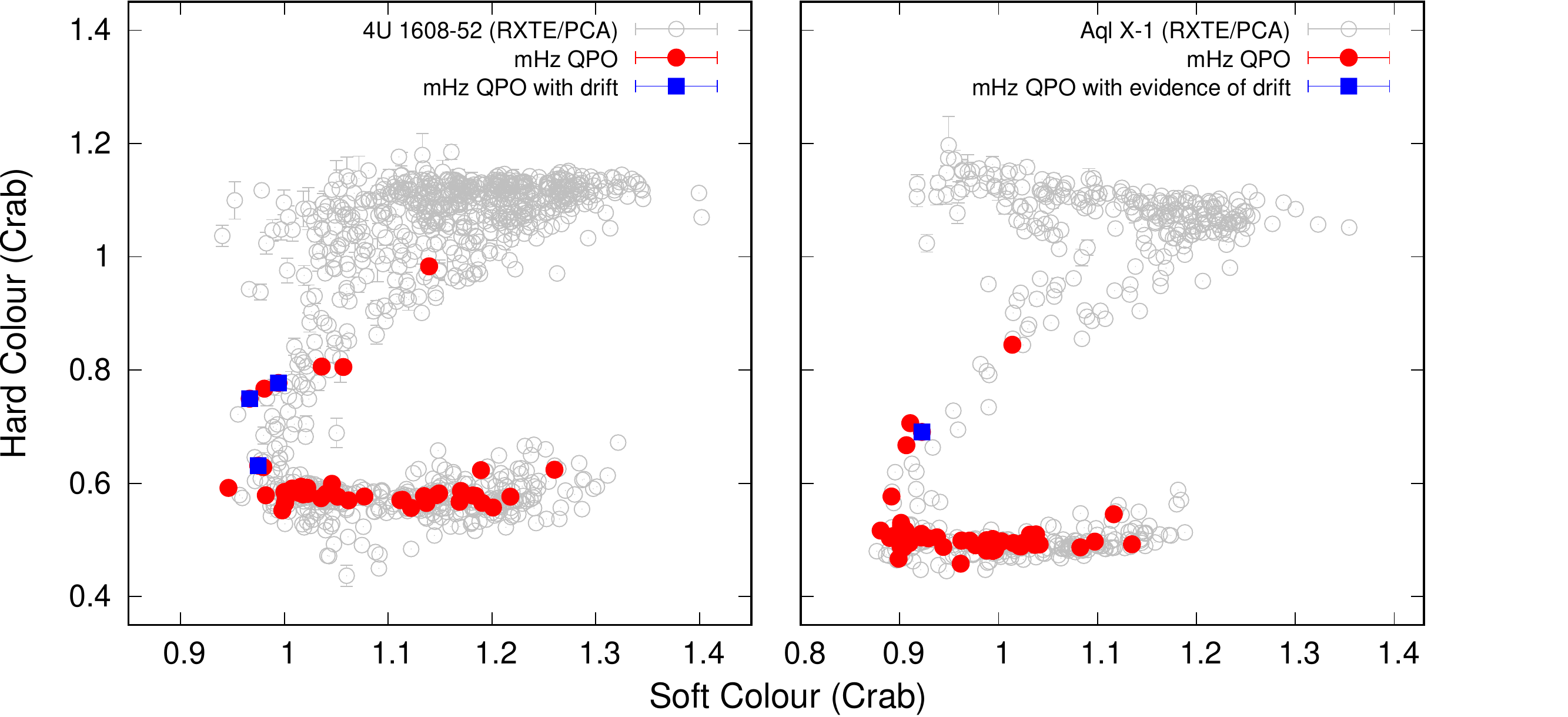}}}
\caption{Colour-colour diagrams (CCDs) of all RXTE observations of 4U~1608--52 (left) and Aql X--1 (right). Each open grey circle corresponds to the averaged colour of the source per RXTE observation, normalised to the colours of the Crab. To calculate the colours, type I X-ray bursts have been removed. Filled red circles mark the location of the source  in the CCD when the mHz QPOs were found. Filled blue squares indicate those cases where we found the mHz QPOs and we observed a frequency drift.}

\label{fig:CCDs}
\end{figure*}

%%%%%%%%%%%%%%%%%%%%%%%%%%%%%%%%%%%%%%%%%%%%%%%%%%%%%%%%%%%%%%%%%%%%%%%%%%%%%%%
%%%%%%%%%%%%%%%%%%%%%%%%%%%%%%%%%%%%%%%%%%%%%%%%%%%%%%%%%%%%%%%%%%%%%%%%%%%%%%%
%%%%%%%%%%%%%%%%%%%%%%%%%%%%%%%%%%%%%%%%%%%%%%%%%%%%%%%%%%%%%%%%%%%%%%%%%%%%%%%
%
%
% Results
%
%
%%%%%%%%%%%%%%%%%%%%%%%%%%%%%%%%%%%%%%%%%%%%%%%%%%%%%%%%%%%%%%%%%%%%%%%%%%%%%%%
%%%%%%%%%%%%%%%%%%%%%%%%%%%%%%%%%%%%%%%%%%%%%%%%%%%%%%%%%%%%%%%%%%%%%%%%%%%%%%%
%%%%%%%%%%%%%%%%%%%%%%%%%%%%%%%%%%%%%%%%%%%%%%%%%%%%%%%%%%%%%%%%%%%%%%%%%%%%%%%
\vspace{-0.5cm}
\section{Results}\label{sec:results}

We detected mHz QPOs in 58 and 57 data segments (in some occasions, the QPO was not present along a whole data segment), in a total of 49 and 47 observations, in 4U 1608--52 and Aql X--1, respectively. In 4U~1608--52, the mHz QPO frequency was between $\sim$ 4.2 and 13.4 mHz, whereas in Aql X--1, the mHz QPO frequency covered between $\sim$ 5.1 and 10.5 mHz. In the lower panels of Figs. \ref{fig:Dynamical1} \& \ref{fig:Dynamical2}, we show examples of X-ray light curves with mHz QPOs in 4U~1608--52.

The CCD of each source is shown in Fig. \ref{fig:CCDs}. The two CCDs are typical of atoll sources \citep{hasinger1989}. Despite the fact that both sources were observed in all spectral states, almost all the mHz QPO detections occurred in the banana branch (BB), i.e., with the softest colours (see filled red circles in Fig. \ref{fig:CCDs}). In a few cases, we observed the oscillations with harder colours, sampling the island state (IS). 
In each source we found one case when the mHz QPOs were detected at hard colour $>0.8$, significantly harder colours than for all other observations with mHz QPOs. In Figs. \ref{fig:Hardest-1608} \& \ref{fig:Hardest-Aql} we show the light curves of those two observations. While in the case of 4U 1608--52 (Fig. \ref{fig:Hardest-1608}) the mHz QPOs are not as clear as those seen in Figs. \ref{fig:Dynamical1} \& \ref{fig:Dynamical2} (even if significant under our assumptions; see Section~\ref{sec:dataanalysis}), the mHz QPOs in Aql X--1 (Fig. \ref{fig:Hardest-Aql}) can clearly be seen in the light curve.

We found that the mHz QPOs disappeared at the time of the occurrence of a type I X-ray burst in 3 cases, 2 in 4U~1608--52 and 1 in Aql X--1. One of our 4U~1608--52 cases was previously reported by \citet{revnivtsev2001}. The other case in this source corresponds to a later observation (obsID 70059-03-01-000; 2002 September). \citet{revnivtsev2001} also reported a QPO peak at 6--7 mHz that disappeared after the onset of a type I X-ray burst in Aql X--1. However, we could not confirm this finding given that the significance of the possible QPO was below the 3$\sigma$ level. 
In those cases in which we detected the mHz oscillations before the burst, but not after it, we found an average rms of $\sim$ 2--3\% and a 3$\sigma$ upper limit of less than $\sim$ 0.5\%, respectively. Note also that, since the length of the segments are always $\gtrsim$ 1 ksec, the minimum frequency we are able to detect is $<$ 1 mHz, and then, even if it is very unlikely, we cannot discard QPOs at frequencies below this limit.
Our results are consistent with those found by \citet{revnivtsev2001} in the same systems and with those found in the remaining sources (\citealt{altamirano2008b}; \citealt{strohmayer2011}; \citealt{lyu2016}; \citealt{Mancuso2019}).

\vspace{-0.25cm}
\subsection{Downward Frequency Drift in 4U~1608--52}

\begin{figure}
\includegraphics[height=5.25cm,width=0.95\columnwidth]{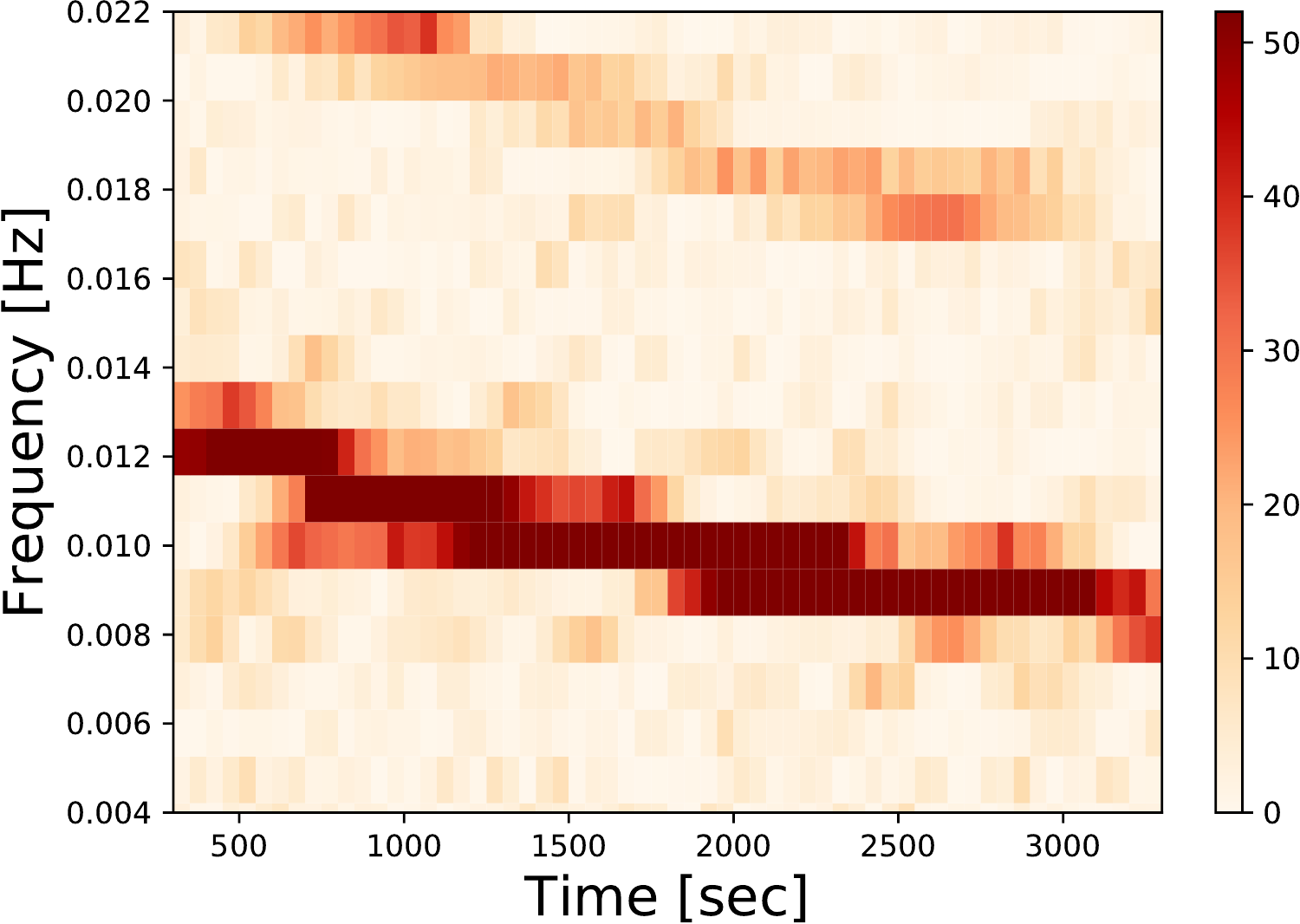}\\
\includegraphics[height=4.75cm,width=0.95\columnwidth]{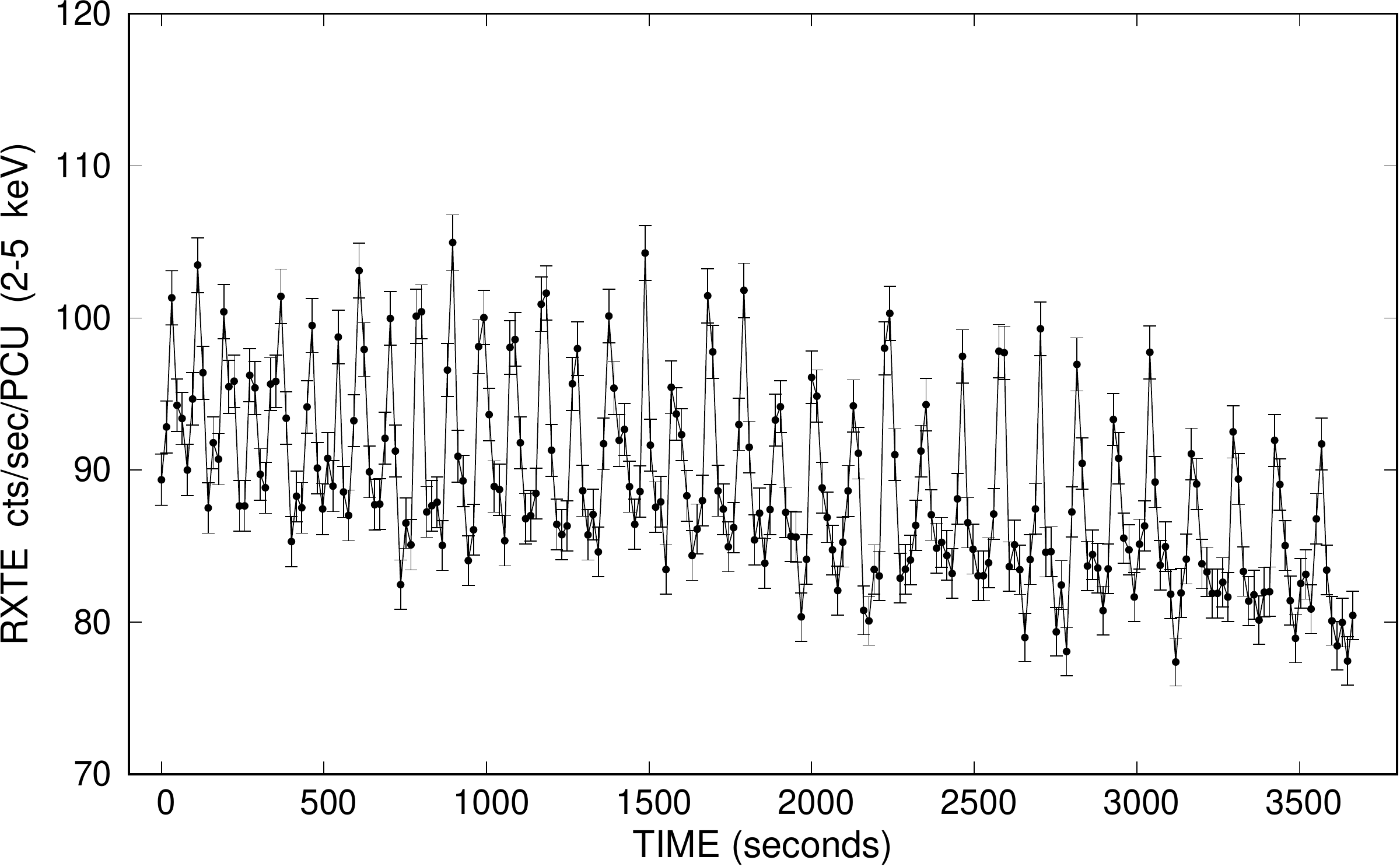}
\centering
\caption{{\it Top:} Dynamical power spectrum (DPS) of the second orbit of the RXTE observation 95334-01-03-06 of 4U~1608--52. To construct the plot, we used a 600-s window sliding with a step of 50 s. The DPS shows that the mHz QPO frequency drifts from $\sim$ 12 mHz down to $\sim$ 8 mHz. Besides the fundamental QPO, a significant ($>3\sigma$) second harmonic is visible. {\it Bottom:} RXTE/PCA background subtracted light curve ($\approx$ 2--5 keV) of the same orbit, with a binning of 16 s. The mHz QPOs are present along the $\sim$ 3.6 ksec long duration.}
\label{fig:Dynamical1}
\end{figure}

\begin{figure}
\includegraphics[height=5.25cm,width=0.99\columnwidth]{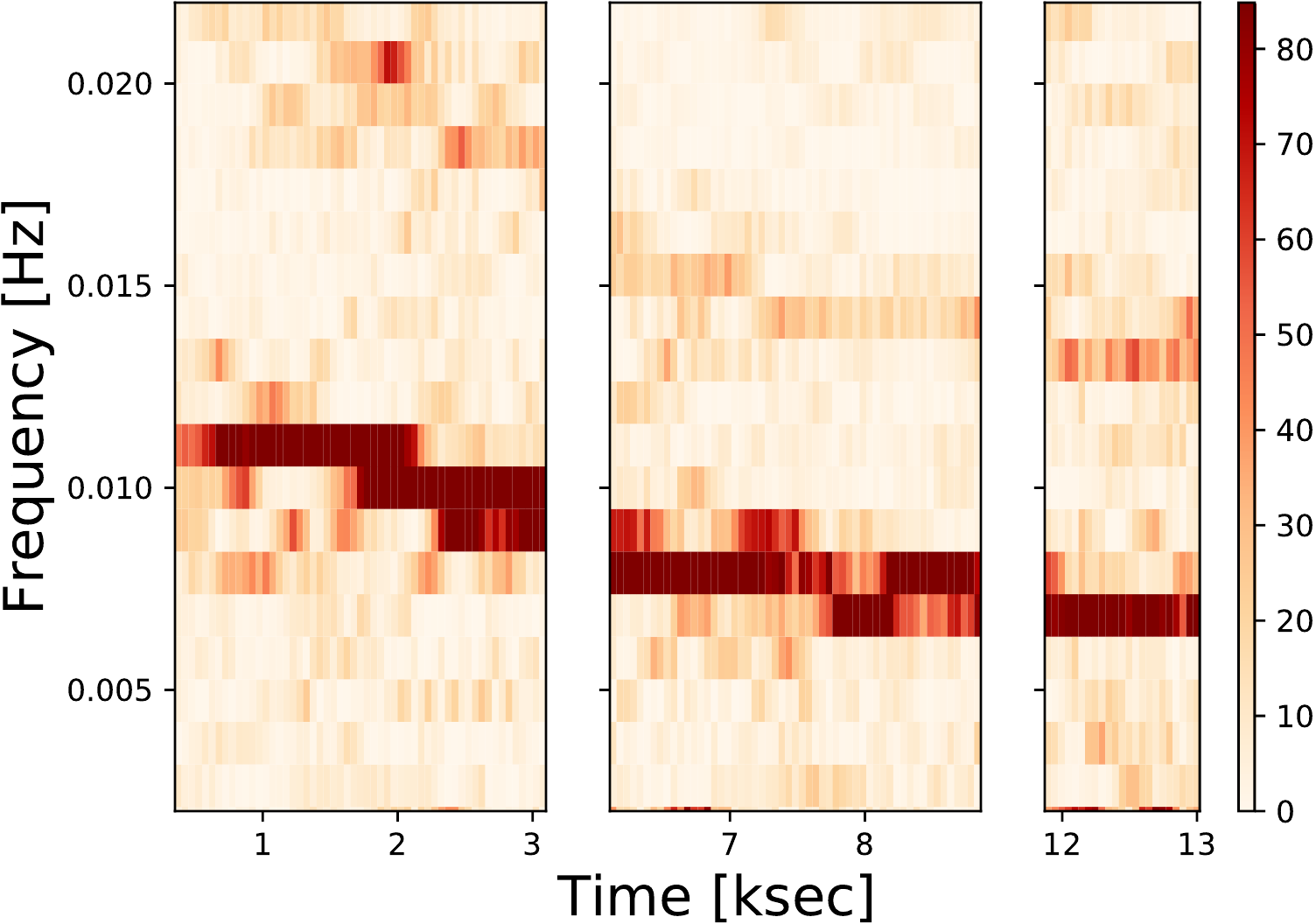}\\
\includegraphics[height=4.75cm,width=0.95\columnwidth]{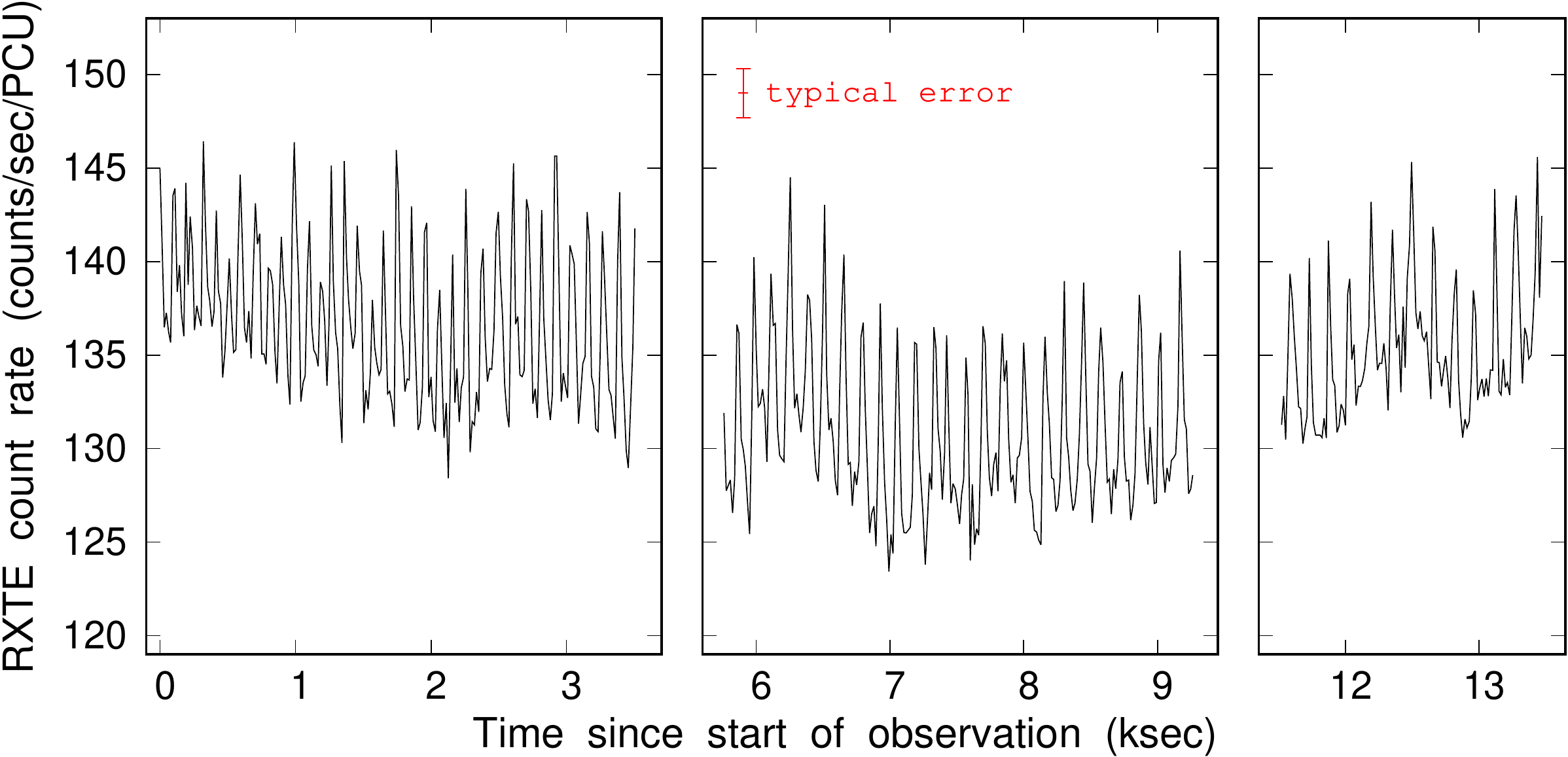}
\centering
\caption{{\it Top:} Dynamical power spectrum (DPS) of the RXTE observation 30062-02-01-01 of 4U~1608--52. To construct the plot, we used a 700-s window sliding with a step of 50 s. The DPS shows that the mHz QPO frequency drifts from $\sim$ 10--11 mHz down to $\sim$ 6--7 mHz. {\it Bottom:} RXTE/PCA background subtracted light curve ($\approx$ 2--5 keV) of the same observation, with a binning of 16 s. The mHz QPOs are present along the three orbits for $\sim$ 13.4 ksec, if we interpolate through the data gaps.}
\label{fig:Dynamical2}
\end{figure}

All data subsets where we detected mHz oscillations have lengths of at most $\sim$ 3.5--3.7 ksec (in both systems). We found frequency drift either within a data segment or by combining contiguous data segments separated by $\sim$ 2.3 ksec data gaps. In 4U~1608--52 we detected 3  cases (that we call group 1) with a significant systematic downward frequency drift. The light curve and the dynamical power spectrum of one of these cases is shown in Fig. \ref{fig:Dynamical1}. To estimate the frequency drift in this observation (obsID 95334-01-03-06; two data segments, with a gap of $\sim$ 2.0 ksec between them), we took the first 900 sec and the last 1300 sec of the second data segment of the observation (the first data segment did not show mHz QPOs), and fitted them with a sinusoidal function plus a constant. The QPO frequency at the beginning of the observation was $\nu$ = $11.74 \pm 0.03$ mHz and at the end of it was $\nu$ = $8.25 \pm 0.02$ mHz. The observation lasted $\sim$ 3.6 ksec and the frequency drifted at an average rate of $\sim$ $0.97$ mHz ksec$^{-1}$. 
Another case is that of obsID 30062-02-01-01 (3 data segments). In this case, we took 1100 sec at the beginning of the first data segment and 1400 sec at the end of the third data segment. The frequency of the oscillations decreased from $\nu$ = $10.17 \pm 0.04$ mHz in the first segment down to $\nu$ = $6.43 \pm 0.03$ mHz in the second one (see Fig. \ref{fig:Dynamical2}). Assuming that the mHz QPOs were present during the data gaps, the drift lasted for about 13.4 ksec, giving an average rate of $\sim$ 0.28 mHz ksec$^{-1}$. 
Figs. \ref{fig:Dynamical1} and \ref{fig:Dynamical2} suggest a correlation between the QPO frequency and the average count rate. However, we note that Fig. \ref{fig:Dynamical2} data leads to a high-scattered correlation (the average intensity decreases and increases while the frequency decreases or remains constant; see Fig. \ref{fig:Dynamical2}).
Given that we only have two cases where we observe this potential correlation, we do not discuss it further (see, e.g., \citealt{altamirano2008b}; these authors found that for 4U~1636--53 the frequency of the mHz QPO is not always correlated to the intensity).
In the last case (obsID 95334-01-04-00), we fitted the same sinusoidal function to the first 750 sec and to the last 1000 sec of the observation. The QPO frequency dropped from $\nu$ = $13.42 \pm 0.08$ mHz to $\nu$ = $10.08 \pm 0.03$ mHz within $\sim$~1.7 ksec, at an average rate of $\sim$ 1.9 mHz ksec$^{-1}$. In all these cases, 4U~1608--52 was in the lower part of the IS, close to the BB (see filled blue squares in Fig. \ref{fig:CCDs}, left). We also found evidence of a downward frequency drift in other 4 cases (group 2). In all these cases, the initial frequency of the QPO was below 9 mHz and decreased by $\lesssim$~1 mHz. These observations sampled the BB, but with soft colour between $\sim$ 1.0 and 1.1 Crab. Finally, we detected stochastic variations of the QPO frequency among data segments in three occasions (group 3). In all these instances, the changes occurred below 9 mHz and within a range of $\lesssim$~2 mHz. In addition, these cases occupied the same portion of the CCD as group 2. All the above results are similar to the findings of \citet{altamirano2008b} in the NS system 4U~1636--53.

Finally, we studied whether the frequency of the mHz QPOs was related to the soft ($\approx$ 2--5 keV) count rate. We found no evidence of a correlation when taking all data available, or only choosing those data where we do not detect frequency drifts. This result was not unexpected, given that the QPO frequency is always below $\sim$ 9 mHz when the source is in the soft state (e.g., similar to that found in 4U~1636--53; \citealt{altamirano2008b}; \citealt{lyu2015}). 
We find evidence of a correlation between the mHz QPO frequency and the soft count rate in the 3 cases where we detected a frequency drift (i.e., only for 4U~1608--52). However, the correlations are different in each case, and therefore it is unclear whether we are seeing a real characteristic of the mHz QPO phenomenon.

\vspace{-0.25cm}
\subsection{Downward Frequency Drift in Aql X--1?}

In Aql X--1 we only found one case with a significant evidence of a downward frequency drift. In this observation (obsID 92034-01-03-00), the frequency of the oscillations was $\nu$ = $9.00 \pm 0.15$ mHz in the first 1000 sec, and decreased to $\nu$ = $6.04 \pm 0.11$ mHz in the last 1000 sec. The mHz QPO frequency drifted at an average rate of $\sim$ 0.92 mHz ksec$^{-1}$ through the $\sim$ 3.2 ksec long duration. Aql X--1 was in the lower part of the IS throughout this observation (see the filled blue square in Fig. \ref{fig:CCDs}, right), consistent with the position in the CCD of 4U~1608--52 during the three cases of frequency drift. We also detected evidence of a downward frequency drift either within or between adjacent data segments (similar to the group 2 of 4U~1608--52), in other 7 occasions. In all these cases we observed the oscillations with a frequency of less than 9 mHz (between $\sim$ 6.1 mHz and $\sim$ 7.5 mHz), and the downward drift  was of the order of $\sim$ 1.3 mHz or less. One out of the seven cases sampled the IS of the CCD (hardest colour); the other 6 cases occurred when Aql X--1 was in the BB.

In addition to the cases described above, we found four cases showing stochastic variations of the QPO frequency ($<$~2 mHz), always when the QPO frequency was below 9 mHz. All these cases, as in the prior group (and similar as groups 2 and 3 of 4U~1608--52), occupied the region of the CCD with the softest colours. Such frequency variations were also observed in 4U~1636--53 (\citealt{altamirano2008b}; \citealt{lyu2015}).

%%%%%%%%%%%%%%%%%%%%%%% GRAFICOS %%%%%%%%%%%%%%%%%%%%%%%%%%%%%%%

%%%%%%%%%%%%%%%%%%%%%%%%%%%%%%%%%%%%%%%%%%%%%%%%%%%%%%%%%%%%%%

%%%%%%%%%%%%%%%%%%%%%%%%%%%%%%%%%%%%%%%%%%%%%%%%%%%%%%%%%%%%%%%%%%%%%%%%%%%%%%%
%%%%%%%%%%%%%%%%%%%%%%%%%%%%%%%%%%%%%%%%%%%%%%%%%%%%%%%%%%%%%%%%%%%%%%%%%%%%%%%
%%%%%%%%%%%%%%%%%%%%%%%%%%%%%%%%%%%%%%%%%%%%%%%%%%%%%%%%%%%%%%%%%%%%%%%%%%%%%%%
%
%
% Discussion
%

%%%%%%%%%%%%%%%%%%%%%%%%%%%%%%%%%%%%%%%%%%%%%%%%%%%%%%%%%%%%%%%%%%%%%%%%%%%%%%%
%%%%%%%%%%%%%%%%%%%%%%%%%%%%%%%%%%%%%%%%%%%%%%%%%%%%%%%%%%%%%%%%%%%%%%%%%%%%%%%
%%%%%%%%%%%%%%%%%%%%%%%%%%%%%%%%%%%%%%%%%%%%%%%%%%%%%%%%%%%%%%%%%%%%%%%%%%%%%%%
\vspace{-0.5cm}
\section{Discussion}

We used nearly 16 years of RXTE archival data to perform the first systematic search and characterisation of the mHz QPOs in the NS LMXBs 4U~1608--52 and Aql X--1. We detected the QPO in the $\sim$ 4.2 to 13.4 mHz frequency range. We found for the first time significant downward frequency drift of the mHz QPOs of 4U~1608--52 in 3 occasions. In the case of Aql X--1, we only found one case with a strong evidence of frequency drift, suggesting that the downward frequency drift might also occur in this system. 

Fig. \ref{fig:CCDs} shows that most part of the mHz oscillations occurred at the softest colours, in the so-called banana branch (BB). Only in a few cases, we detected the QPOs at harder colours, in the island state (IS) or at the transition between the BB and the IS. We also found that the oscillations disappeared after the onset of a type I X-ray burst, twice in 4U~1608--52 (one of them previously reported by \citealt{revnivtsev2001}) and once in Aql X--1. All these characteristics are consistent with those found in the rest of the systems with mHz QPOs (\citealt{altamirano2008b}; \citealt{strohmayer2011}; \citealt{strohmayer2018}; \citealt{Mancuso2019}), suggesting that these mHz QPOs are of the same type as those originally discovered by \citet{revnivtsev2001}. Therefore, assuming that the current interpretation of (marginally stable) nuclear burning is correct, then the mHz QPOs observed in 4U~1608--52 and Aql X--1 are the result of He burning on the NS surface \citep[e.g.,][]{heger2007,keek2009}.

Up to now, only two sources had shown a decrease of the QPO frequency with time (\citealt{altamirano2008b}; \citealt{Mancuso2019}). \citet{altamirano2008b} found a systematic frequency drift of the QPOs in 4U~1636--53 at an average rate between 0.07 and 0.15 mHz ksec$^{-1}$. The average rate of decreasing measured by \citet{Mancuso2019} in EXO~0748--676 was of 0.26 and 0.56 mHz ksec$^{-1}$ in two separate occasions. In our three instances of frequency drift detected in 4U~1608--52, we estimated a decrease at an average rate of $\sim$ 0.3, $\sim$ 1.0 and $\sim$ 1.9 mHz ksec$^{-1}$.\footnote{In 4U~1636--53, EXO~0748--676 and 4U~1608--52, the average rate was calculated, when needed, interpolating through data gaps.} In particular, these last two values are the fastest average rate measured for a frequency drift. However, we note that these average rates depend on the length of the observations. 
If we take as an example the observation shown in Fig. \ref{fig:Dynamical2}, we would have measured different frequency drifts if we had only used one or two of the 3 data segments (e.g., no drift if we had used the 3rd data segment).
The fact that the observation was long enough while the mHz QPOs were drifting made possible to estimate an average rate. In the same way, faster rates within a data segment might have occurred in any of the sources where the drift has been observed.

We also found that the highest values of the rate correspond to those two observations with the hardest colours. Although this is a result based on only three cases, it supports the suggestion by \citet{Mancuso2019} that the rate at which the frequency drops decreases as the source becomes softer. 

In the case of Aql X--1, we only found one observation with strong evidence of a downward in the mHz QPO frequency. We calculated an average drift rate of $\sim$ 0.9 mHz ksec$^{-1}$, consistent to one of the 4U~1608--52 cases, and higher than those reported in the previous sources. Although the downward drift is significant and the source is in the intermediate state (as expected based on the results on 4U~1636--53, 4U~1608--52 and EXO~0748--676), we consider this case only as a strong evidence as the QPO has been seen to vary stochastically when the average frequency is below 9 mHz (e.g., this work and \citealt{altamirano2008b}).

It is possible that the reason we did not detect more significant downward frequency drifts in Aql X--1 is that we were unlucky with the observational sampling of Aql X--1 intermediate states. However, the explanation is probably related to the length of the data segments: RXTE monitored Aql X--1 many times with single-orbit snapshots per day, most of them with integration times lower than 2.0 ksec (and only few with integration times as long as $\sim 3.5$ ksec). Aql X--1 data are, at most, similar to the cases of data segments two and three of the observation of 4U~1608--52 shown in of Fig. \ref{fig:Dynamical2}.

%%%%%%%%%%%%%%%%%%%%%%% GRAFICO %%%%%%%%%%%%%%%%%%%%%%%%%%%%%%%
\begin{figure}
\includegraphics[height=5.30cm,width=0.99\columnwidth]{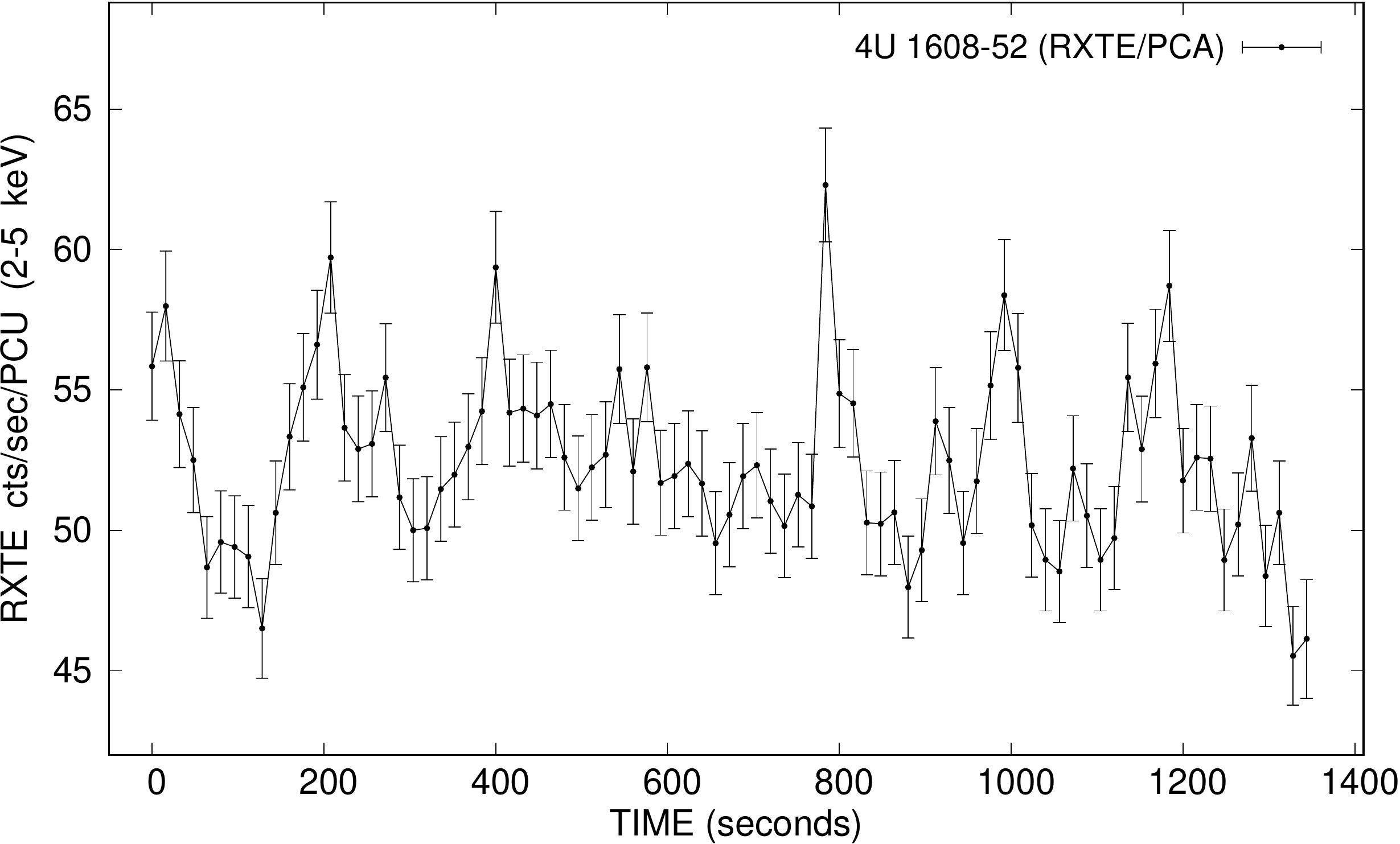}
\centering
\caption{Background-subtracted 16 s binned light curve ($\approx$ 2--5 keV) of the observation with the largest hard colour that shows mHz QPOs of 4U~1608--52 taken with RXTE/PCA.}
\label{fig:Hardest-1608}
\end{figure}
%%%%%%%%%%%%%%%%%%

\vspace{-0.25cm}
\subsection{mHz QPOs and accretion rate}

In atoll sources, the position of a system in the CCD can be parametrised by the length, $S_{a}$, of the curve along the path traced by the source in that diagram (see, e.g., Fig. 1 in \citealt{mendez1999}). This length is commonly normalised, taking the value $S_{a} = 1$ at the top right-hand vertex, and $S_{a} = 2$ at the bottom left-hand vertex of the CCD. 
\citet{mendez1999} found that in 4U~1608--52 the kHz QPO frequency increased while the system moved from the island to the banana state. The fact that the kHz QPO frequency and $S_{a}$ were well correlated, while there was no single correlation between the frequency and the count rate when the source was observed at different luminosities, led the authors to suggest that $S_{a}$ is a good proxy for $\dot{M}$ (see also \citealt{hasinger1989}). 

\citet{altamirano2008b} found that, in 4U~1636--53, all the mHz oscillations occurred when the system was in either the BB, or in the transition between this state and the IS. We find something similar in both 4U~1608--52 and Aql X--1. However, we also found two cases (one per source, see Figs. \ref{fig:Hardest-1608} \& \ref{fig:Hardest-Aql}), in which mHz QPOs are detected in the upper part of the IS,\footnote{This is similar to what it was found by \citet{Mancuso2019} in EXO~0748--676; see their Fig. 2; however we  note that EXO 0748--676 colours are affected by the high inclination of the system.} i.e., at lower values of $S_a$.
If we assume that the relation between $S_{a}$ and the mass accretion rate is correct (see, e.g., \citealt{hasinger1989}; \citealt{mendez1999}; \citealt{zhang2011}), then our results show that the marginally stable burning occurs in a larger range of mass accretion rate than models predict \citep[e.g.,][]{heger2007}. In fact, for 4U~1608--52, and assuming a distance of 3.7 kpc, we found that the luminosity when the source was in the upper part of the IS was $\sim 2.2 \times 10^{36}$ erg s$^{-1}$ in the 3--20 keV energy range, i.e., an order of magnitude lower than the highest luminosity observed by \citet{revnivtsev2001}.
We note that if \citet{heger2007} model is correct, then our results would indicate that the local accretion rate onto the NS is always close to the Eddington limit, even if we see significant changes in the source spectra (parametrised by the position in the CCD) and average X-ray luminosity.

\vspace{-0.25cm}
\subsection{mHz QPO frequency drift and the NS crust properties}
%
%
%%%%%%%%%%%%%%%%%%%%%%% GRAFICO %%%%%%%%%%%%%%%%%%%%%%%%%%%%%%%
\begin{figure}
\includegraphics[height=5.30cm,width=0.99\columnwidth]{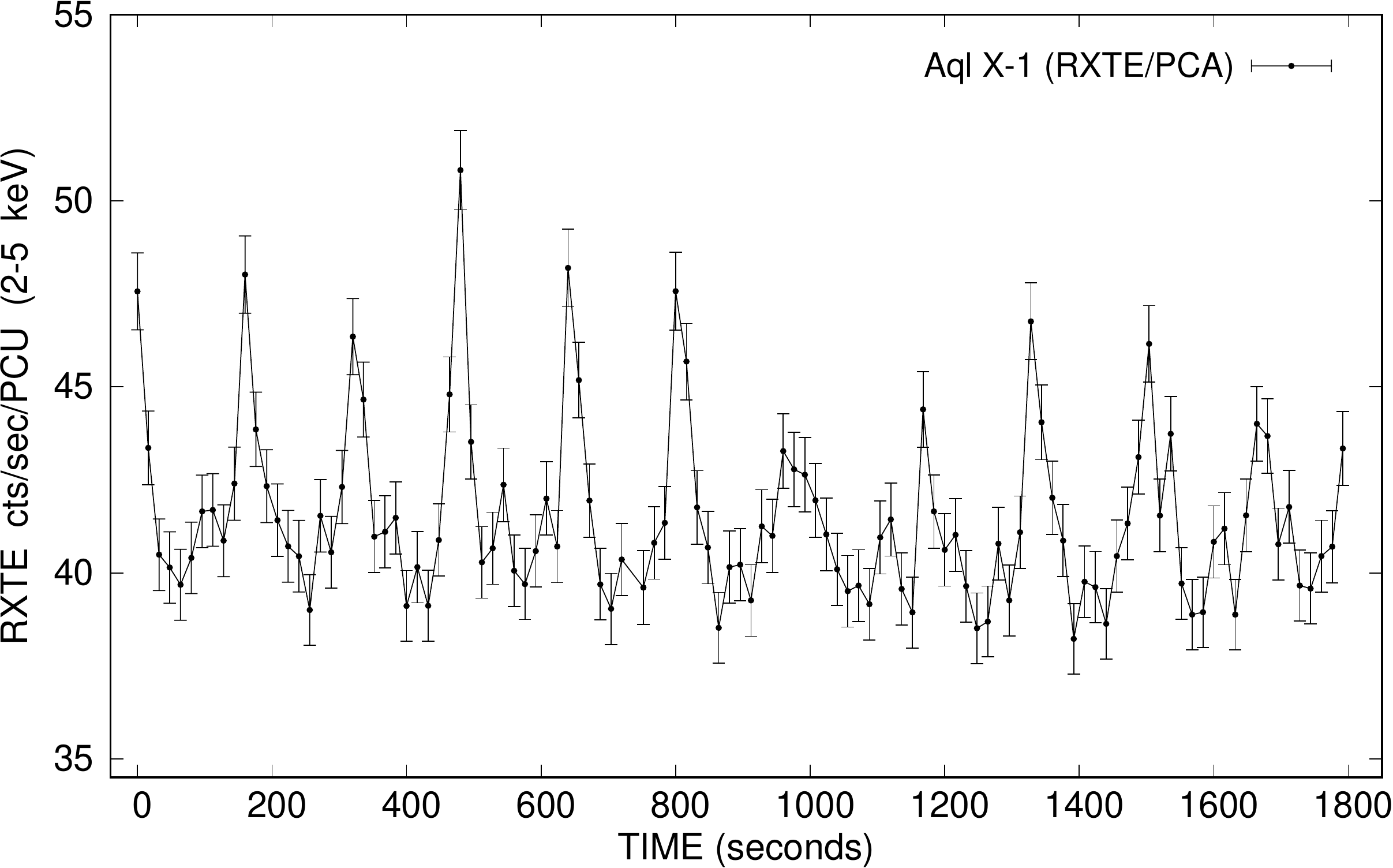}
\centering
\caption{Background-subtracted 16 s binned light curve ($\approx$ 2--5 keV) of the observation with the largest hard colour that shows mHz QPOs of Aql X--1 taken with RXTE/PCA.}
\label{fig:Hardest-Aql}
\end{figure}
%%%%%%%%%%%%%%%%%%%%%%%%%
%
\citet{altamirano2008b} discovered that, when 4U~1636--53 was near to the transition between the IS and the BB, the mHz QPO frequency decreased systematically with time (see also \citealt{lyu2014,lyu2015}), and once it dropped below 9 mHz, an X-ray burst took place within a few kiloseconds afterwards. The case of EXO 0748--676 is similar, as \citet{Mancuso2019} found two instances of frequency drift, in an area of the CCD that the authors identified as close to the intermediate state. We found a similar result in 4U~1608--52: the three cases where we found significant downward frequency drift occurred at a time when the source was near the transition between the soft and hard states. 
The fact that these frequency drifts are detected roughly in the same part of the CCD independently of the source (4U~1636--56, EXO~0748--676, and 4U~1608--52; our Aql X--1 results are still consistent with this picture) strongly suggests a relation between the source state (which is likely set by the mass accretion rate) and the expected QPO frequency evolution. 

\citet{keek2009} showed that the QPO frequency decreases with time if the heat flux from the crust decreases, i.e., the drift could be the result of the cooling of deep layers. However, \citet{keek2009} did not discuss why the cooling of the deep layers could change depending on the source state. We speculate that the change of behaviour might be related to the change in accretion rate.
Assuming that the accretion rate is an increasing function of the parameter $S_{a}$ (see above), as the source changes its spectral state, from the IS to the BB, the mass accretion rate increases, as well as the luminosity. This implies, in turn, an increase of the temperature of the layer where the burning takes place. %((a) see eq (3) of Heger?; (b) importa ya sea por $\epsilon$ o y (segun la misma ecuacion)). 
Consequently, this increasing in the temperature balances the cooling of the deeper layers, until it becomes high enough to compensate completely the cooling, and therefore no drift would occur at all. We speculate that this would happen near the bottom left-hand vertex of the CCD, i.e., at a mass accretion rate compatible with $S_{a} \simeq 2$. 

\citet{altamirano2008b} found that the QPO frequency drifted always with an observed initial frequency between 10.7 and 14.3 mHz.
\citet{Mancuso2019} also observed a starting frequency of the QPO of $\sim$ 13 mHz in one of their cases.\footnote{In the another case, the authors {\it began} to observe the QPO frequency at $\sim$ 8.2 mHz, which is below 9 mHz, but it is not inconsistent with a possible onset of the QPO frequency above 9 mHz.}
In 4U~1608--52, we detected a similar result: an initial frequency above 9 mHz, between 10.2 and 13.4 mHz. 
Our findings, combined with the previous ones, suggest that the onset of the QPOs that show a downward drift is at a frequency $\gtrsim$ 10 mHz. Moreover, given the relatively wide range of initial QPO frequency and the different hard-colour values where the downward frequency drift was observed, a relation between the initial mHz QPO frequency drift and the spectral state cannot be excluded.

\vspace{-0.5cm}
\section{Summary}\label{sec:summary}
Previous works showed that mHz QPOs with systematic frequency drifts occur in the NS LMXBs 4U~1636--53 and EXO~0748--676.
Here we discovered a significant drift of the frequency of the mHz QPOs in 4U~1608--52, and strong evidence of such drift in the mHz QPOs observed in Aql X--1. 
Our results are consistent with previous findings in other systems, and strongly suggest that the mHz QPOs can undergo downwards frequency drifts if they are observed when the systems in which they occur is in the intermediate spectral state.

%%%%%%%%%%%%%%%%%%%%%%%%%%%%%%%%%%%%%%%%%%%%%%%%%%%%%%%%%%%%%%%%%%%%%%%%%%%%%%%
%%%%%%%%%%%%%%%%%%%%%%%%%%%%%%%%%%%%%%%%%%%%%%%%%%%%%%%%%%%%%%%%%%%%%%%%%%%%%%%
%%%%%%%%%%%%%%%%%%%%%%%%%%%%%%%%%%%%%%%%%%%%%%%%%%%%%%%%%%%%%%%%%%%%%%%%%%%%%%%
%
%
% Summary 
%
%
%%%%%%%%%%%%%%%%%%%%%%%%%%%%%%%%%%%%%%%%%%%%%%%%%%%%%%%%%%%%%%%%%%%%%%%%%%%%%%%
%%%%%%%%%%%%%%%%%%%%%%%%%%%%%%%%%%%%%%%%%%%%%%%%%%%%%%%%%%%%%%%%%%%%%%%%%%%%%%%
%%%%%%%%%%%%%%%%%%%%%%%%%%%%%%%%%%%%%%%%%%%%%%%%%%%%%%%%%%%%%%%%%%%%%%%%%%%%%%%

\vspace{-0.25cm}
\section*{Data availability}

The data underlying this article are publicly available in the High Energy Astrophysics Science Archive Research Center (HEASARC) at \url{https://heasarc.gsfc.nasa.gov/db-perl/W3Browse/w3browse.pl}.

\vspace{-0.3cm}
\section*{Acknowledgments}
GCM acknowledges support from the Royal Society International Exchanges ``the first step for High-Energy Astrophysics relations between Argentina and the UK''. 
DA acknowledges support from the Royal Society. JAC and GCM were partially supported by PIP 0102 (CONICET). This work received financial support from PICT-2017-2865 (ANPCyT). 
Lyu is supported by National Natural Science Foundation of China (grant No.11803025), and Hunan Provincial Natural Science Foundation (grant No. 2018JJ3483). JAC was also supported by grant PID2019-105510GB-C32/AEI/10.13039/501100011033 from the Agencia Estatal de Investigaci\'on of the Spanish Ministerio de Ciencia, Innovaci\'on y Universidades, and by Consejer\'ia de Econom\'ia, Innovaci\'on, Ciencia y Empleo of Junta de Andaluc\'ia as research group FQM-322, as well as FEDER funds.
This research has made use of data and/or software provided by the High Energy Astrophysics Science Archive Research Center (HEASARC), which is a service of the Astrophysics Science Division at NASA/GSFC and the High Energy Astrophysics Division of the Smithsonian Astrophysical Observatory. This research has made use of NASA's Astrophysics Data System.

\vspace{-0.25cm}
%\begin{thebibliography}{mn2e}
\bibliographystyle{mnras}
\bibliography{biblio2.bib}
%\end{thebibliography}

%\bibliographystyle{aa}
\vspace{5mm}

%\bibliography{../../../Most_complete_bib}
%\begin{thebibliography}

\label{lastpage}
\end{document}